\documentclass[12pt]{article}
\usepackage{amssymb}
\usepackage{graphicx}
\usepackage{indentfirst}
\usepackage[square, super, comma, sort&compress]{natbib} 
\usepackage[a4paper,left=2.5cm, right=2.5cm, top=2cm, bottom=3cm]{geometry}
\usepackage{titlesec}
\titleformat*{\section}{\large\bfseries}

\usepackage[utf8]{inputenc}
\usepackage{url}
\usepackage{hyperref}
\usepackage{amsmath}
\usepackage{amsfonts}
\usepackage{float}
\usepackage{lipsum}
\usepackage{xcolor}
\usepackage{natbib}
\usepackage{caption}
\usepackage{multirow}

\begin{document}


\begin{center}

	{\Large \textbf{Recent development of optical}}\\ \vspace{1em}
	{\Large \textbf{electric current transformer and its obstacles}}\\ \vspace{3em}

	{\large Yu-Xuan Chen, Jing Sun, Bo-Qi Meng*}\\ \vspace{3em}

\end{center}


\begin{center}
    \centering
	\rule{150mm}{0.5mm}
\end{center}

\begin{abstract}

Conventional electromagnetic induction-based current transformers suffer from issues such as bulky and complex structures, slow response times, and low safety levels. Consequently, researchers have explored combining various sensing technologies with optical fibers to develop optical current transformers that could become the primary choice for power systems in the future. With the maturation of optoelectronic technology, optical current transformers have emerged. They offer outstanding advantages, including high sensitivity, integration, stability, and the ability to operate in complex environments. This review categorizes optical current transformers based on different principles, including all-fiber current transformers, those based on magnetostrictive effects, magneto-optic effects, and thermal effects. It also discusses their principles, structures, manufacturing techniques, and signal processing, while forecasting their future development trends.

\end{abstract}

\begin{center}
    \centering
	\rule{150mm}{0.5mm}
\end{center}

\vspace{5mm}


\section{Introduction}
Traditional electromagnetic induction-based current transformers are no longer able to meet the increasing demands of power systems due to issues such as poor insulation performance, excessive electromagnetic interference, and large size and weight \cite{wu2020recent, bohnert2002temperature}. In contrast, optical current transformers offer stronger corrosion resistance, superior insulation performance, greater immunity to interference, easier installation and maintenance, and significantly reduced costs. These advantages have drawn significant attention to optical current transformers, positioning them as an important development trend in current sensing technology.
Research on optical current transformers began in the 1960s. Several research institutions, companies, and universities in the United States and Japan devoted substantial efforts to optical current transformers and achieved promising results \cite{sima2022improving,zhao2020design}. Subsequently, the United Kingdom and Germany also transitioned to practical applications, deploying optical current transformers in grid operations and collecting experimental data \cite{lenner2019long,elsherif2022optical}. In comparison, China started its research on current transformers relatively late and faced technological challenges. However, with rapid economic development, Chinese schools, institutions, and enterprises have actively engaged in optical current transformer research. Notably, both Tsinghua University and Huazhong University of Science and Technology have made significant contributions. Since the birth of the first optical current transformer in 1963, these devices have experienced rapid development. Increasing numbers of scholars have dedicated themselves to optical current transformer research, leading to diverse applications across various fields. As technology matures, achieving higher sensitivity and precision in current measurement remains a key focus. Optical current transformers overcome the limitations of traditional designs, such as bulky and complex structures and low safety levels \cite{nasir2021design,bohnert2007fiber}. They align well with trends in microprocessor protection, digital energy metering, and intelligent automation, offering broad prospects for practical applications.

\section{All-fiber optic current transformer}
\subsection{Sagnac type}
The Sagnac interferometric fiber optic current transformer is derived from fiber optic gyroscope technology \cite{wang2001fiber, yu1994practical, sasaki2015temperature}. It is a circular interferometer based on the Sagnac effect. Due to its advantages of full solid state, high reliability, flexible configuration, long life, fast startup and fast response, this technology has attracted much attention, developed rapidly and has been widely used \cite{mihailovic2021fiber}. After the light beam emitted by the light source passes through the fiber polarizer, the linearly polarized light is divided into two beams with the same polarization state, which are transmitted in opposite directions of the polarization-maintaining fiber. It is converted into circularly polarized light through a /4 wave plate, and one of the beams passes through a delay device before being converted into circularly polarized light through a /4 wave plate. At this time, the two beams have the same rotation direction but opposite directions. The circularly polarized light in the sensor coil rotates under the influence of the Faraday effect, converges at one point to interfere, and finally the signal is demodulated by signal processing to obtain the current size.
In the late 1980s, P A Nicati and Ph Robert initially constructed a fiber optic current transformer based on a stable Sagnac structure, and tested it and found that the dynamic measurement range of this type of current transformer was greatly increased. Since then, this structure has been widely popularized and developed. Domestic and foreign scholars have mainly made improvements on linear birefringence, circular birefringence, temperature, vibration and other issues, and have made great progress. In 1994, A Yu and AS Siddiqui proposed a new practical Sagnac interferometric fiber optic current transformer using a 3×3 fiber coupler, and conducted a detailed analysis of the two main problems of linear birefringence caused by bending and circular birefringence caused by twisting. Compared with traditional electromagnetic current transformers, this new type of current transformer is more temperature stable. In 1999, Hermann Lin proposed a Sagnac interferometer current transformer that is not affected by environmental vibration. The demodulation circuit of the transformer is simplified by using a new passive demodulation technology, which can achieve a measurement sensitivity of about 4.5 $\mu$rad/(Arms turns) in both static and dynamic environments, and the average distortion rate is always less than 0.9\%. In 2004, Masao Takahashi of Japan Aviation Electronics Industry Co., Ltd. proposed a Sagnac interferometer type fiber optic current transformer using a single-mode optical fiber lead-in line. By using a depolarizer in the Sagnac coil, the occurrence of non-reciprocal phase error is reduced. This configuration makes the current transformer cheaper to manufacture, with higher measurement accuracy, and the temperature characteristics meet the ratio error within 0.2\%. In 2014, Han Chunyang of Xi`an Jiaotong University proposed a single-mode microfiber Sagnac ring interferometer based on a wavelength-scale diameter single-mode microfiber \cite{lin1999modified}. The sensitivity of the structure to the environmental refractive index and temperature reaches 12500 nm/RIU and 3pm/℃ respectively.

At present, the fiber optic current transformer based on the Sagnac structure has attracted much attention for its high sensitivity. Although researchers have conducted a large number of innovative experiments, high sensitivity and large-range measurement cannot be achieved at the same time. In view of this, in 2023, Liu Chaoyi of Yanshan University proposed a fiber Sagnac interferometer with polarization-maintaining fiber and fiber Bragg grating in the loop [14]. The polarization-maintaining fiber and fiber Bragg grating in the loop are designed as a fiber Sagnac interferometer. Due to the limitation of the free spectrum range, the polarization-maintaining fiber can show high strain measurement sensitivity and limited measurement range. In contrast, the fiber Bragg grating has lower strain sensitivity but a larger measurement range. By combining the polarization-maintaining fiber and the fiber Bragg grating in the Sagnac interferometer, compared with the measurement range of 0~1058.9$\mu\epsilon$ and the sensitivity of 15.11 pm/$\mu\epsilon$ of the ordinary Sagnac interferometer, the measurement range of the transformer is increased to 0~5187$\mu\epsilon$, and the average sensitivity is as high as 17.08 pm/$\mu\epsilon$. This method provides a new idea for measuring strain transformers with large range and high sensitivity.

\subsection{Reflective type}
The reflective fiber optic current transformer is essentially a light wave polarization interferometer \cite{bohnert2019polarimetric, muller2016inherent, peng2013fiber}. The two beams reflected by the laser at the end of the fiber optic sensor coil propagate alternately, instead of using two counter-propagation modes of the same polarization like the Sagnac interferometer. This is a common optical path structure based on single optical path propagation. A beam of light is emitted by the light source and transmitted to the polarizer through the coupler. The polarizer converts the beam into linear polarized light. After the beam is fused, it is decomposed into two linear polarized lights in orthogonal directions. After passing through the phase modulator wave plate, it is converted into two circularly polarized lights, one right-handed and one left-handed. In the sensing fiber ring, it is affected by the Faraday magneto-optical effect, causing a phase difference between the two circularly polarized lights. After the two circularly polarized lights are reflected by the reflector, the left-handed light turns into the right-handed light and the right-handed light turns into the left-handed light. Then, it is affected by the Faraday effect and a phase difference is generated. When the circularly polarized light beam passes through the /4 wave plate again, it is converted into linearly polarized light, which interferes in the polarizer and is then transmitted to the signal processing system through the coupler, thereby obtaining the phase difference between the two polarized lights and indirectly measuring the current.

In 1994, Swiss scholar Guido Frosio first proposed a new all-fiber current transformer, a reciprocal fiber reflection interferometer that measures current through the Faraday effect. The unnecessary reciprocity effect caused by residual birefringence in the optical fiber can be eliminated to a large extent. Canadian scholar Zhou Sheng proposed a new technology to reduce bending-induced linear birefringence in a reflective fiber-optic current transformer through a fiber polarization rotator. The measurement error of the fiber polarization rotator solution is only 0.06\%, while the measurement error of the non-fiber polarization rotator solution is 1.2\%, which provides an order of magnitude improvement. In 2020, Xie Xiaojun of the Hunan Institute of Metrology and Testing proposed a reflective Sagnac-type fiber-optic current transformer, produced a prototype, and conducted measurement experiments on the prototype. The test results show that the measurement accuracy is higher than ±0.2\%, and the ratio error under temperature, vibration, and magnetic field changes is less than ±0.2\%, and the measurement accuracy reaches 0.2 level. Furthermore, reflective-type FOCS is combined with vibration sensor \cite{yu2022polarimetric, yu2023simultaneous}, which can provide warning when the fiber link is being attacked.

\subsection{Current Transformer Based on Magneto-Optical Effect}

As early as 1845, M. Faraday discovered the Faraday magneto-optical effect, and in 1986, the measurement of current based on the Faraday magneto-optical effect was initially realized. Polarized light parallel to the direction of the magnetic field is obtained through a polarizer, and when the polarized light passes through a medium, it is deflected, and the deflection angle $\theta$ is obtained by the analyzer. This phenomenon of polarized light rotating due to a magnetic field is called the Faraday effect, which can also be called the magneto-optical rotation effect \cite{frosio1994reciprocal, chu1992faraday}.

The Verdet constant was discovered by E. Verdet in 1854, and in-depth research was conducted based on the phenomenon of magneto-optical effect. Experimental analysis shows that this phenomenon is related to the constant of the material itself, which is called the Verdet constant. The Verdet constant is very weak in most substances, but it will increase in glass doped with rare earth ions. During the current measurement process of the magneto-optical current transformer, the optical glass sensor head will not have charge retention or short-term discharge due to faults in the measured line (such as open circuit or short circuit), so it has the same measurement performance under transient and steady-state conditions. The measurement accuracy and stability of this type of transformer are affected by many factors. Existing research mainly analyzes the impact of linear birefringence and temperature on its performance. In actual situations, the performance of the magneto-optical current transformer is mainly affected by the combined factors of the fluctuation of the wavelength of the incident light and the accuracy of the manufacturing and assembly process (such as the angle error of the transmission axis of the polarizer and the analyzer). The main sensing part of the magneto-optical current transformer is highly integrated with micro-optical components. The tiny structure encapsulates a series of miniature optical devices such as magneto-optical crystals, collimators, polarizers, etc. It has the characteristics of small size, light weight, and reliable structure, and can be easily used in a variety of occasions.

Thanks to the above advantages, this structure has received widespread attention. The three most common magneto-optical materials are glass, crystal and ceramic materials. Magneto-optical glass has disadvantages such as high absorption rate and low thermal conductivity, while magneto-optical ceramic materials have the advantage of simple preparation process. In contrast, magneto-optical crystals have low absorption coefficient and large Verdet constant, and have received higher attention. In 2022, Liu Jiteng proposed a magneto-optical current transformer using TGG crystal as the sensing material [20], which uses the advantage of the high Verdet constant of TGG crystal to improve the sensitivity of the optical current transformer. Its sensing coefficient for DC measurement is 0.0683±0.0002, and its sensing coefficient for industrial frequency AC measurement is 0.0681. The relative error is less than 0.58\%. Compared with the terbium aluminum borosilicate current transformer (sensing coefficient is 0.0124±0.0005), it has higher measurement accuracy and a sensitivity increase of 40\%.

\subsection{Current Transformer Based on Magnetostrictive Effect}
The magnetostrictive effect was first discovered by J.P. Joule in 1842. The principle of the magnetostrictive effect that is, when a magnetostrictive material is subjected to a magnetic field, the material's shape exhibits a deformation phenomenon \cite{liu2021applications, lopez2019fiber, lopez2019fiber2}. When the magnetic field disappears, the material returns to its original shape. The magnetostrictive effect can be divided into two types: bulk magnetostriction and linear magnetostriction. Bulk magnetostriction refers to the change in volume, and linear magnetostriction refers to the change in length. In 1865, E. Villari observed that when stress is applied, magnetostrictive materials expand and contract due to the magnetostrictive effect. The magnetostrictive coefficient can most intuitively reflect the magnetostrictive effect. What J.P. Joule discovered was actually the negative magnetostrictive effect, that is, as the magnetic field increases, the length of the magnetostrictive material decreases. However, the positive magnetostrictive effect was also discovered. When a material has a positive magnetostrictive effect, it is called the positive magnetostriction coefficient. Usually, the magnetostriction coefficient is an inherent property of the material, which is related to the composition, crystal structure and other properties of the material \cite{chen2023review}. According to its principle, it can be divided into two categories: interferometric fiber optic current transformer and fiber grating current transformer.

With the development of fiber optic sensing technology, some progress has been made in the research on the sensitivity of fiber grating current transformers. Generally speaking, there are two ways to improve the sensitivity: one is to improve the demodulation system of the fiber grating current transformer, and the other is to improve the main body of the fiber grating current transformer. The improved demodulation method is the most widely used. The demodulation methods of fiber grating include sideband demodulation method, interference demodulation method and matching demodulation method. In 2021, Ma Yuehui of Shanghai University proposed a fiber grating current transformer based on magnetostrictive effect. The fiber grating is fixed on the magnetostrictive material, and the current is realized by the influence of the magnetic field generated by the energized solenoid. Zhou Minghui from Huazhong UST has also offered a promising shceme for demodulating the current and temperature\cite{zhou2021simultaneous}. The FBG current transformer has achieved a maximum Bragg wavelength drift of 0.773 nm in the current range of 0~5 A, and a sensing sensitivity of up to 0.184 nm/A. The structures of optical materials and magnetostrictive materials are mostly realized by physical means such as pasting or winding. Therefore, the development trend of such transformers in the future should be to aim at high integration, high sensitivity and combined structural processing, and strive to improve the performance of magnetostrictive fiber Bragg grating transformers.

\subsection{Current Transformer Based on Thermal Effect}
The sensing principle of the current transformer based on thermal effect is that the refractive index of the optical fiber material changes with temperature \cite{lusher2001current, strumpler1996polymer, meijer1986thermal, hammerschmidt2006new}. This type of transformer initially used metal or thermistor. When the temperature changes, the length and refractive index of the micro-nano optical fiber change, resulting in a shift in the resonant wavelength. In essence, they all use the optical fiber temperature sensing principle. Current transformers based on thermal effect are divided into two types: grating optical fiber and micro-nano optical fiber. Micro-nano optical fiber technology is developing rapidly. The micro-nano optical fiber is wrapped around the thermistor material, and current is applied to the thermistor material to generate Joule heat to increase the metal temperature, causing the refractive index of the micro-nano optical fiber to change, and the current size is obtained. Simple principle, high measurement accuracy, and compact structure are the advantages of this type of current transformer. However, micro-nano optical fiber is easily contaminated by air, complicated to operate, and easy to break. Therefore, new micro-nano optical fiber structures and suitable materials are one of its main development directions in the future.

\section{Conclusion}
This table compare the performance of various types of optical current transformers. Optical current transformers overcome the weaknesses of traditional electromagnetic current transformers, and conform to the development trend of microcomputer protection, digitalization, intelligence, and automation of electric energy metering, and have broad development prospects. Even after years of research, fiber optic current transformer technology has made great strides, but it still cannot be applied on a large scale. 

\begin{table}[h]
\centering
\begin{scriptsize}
\begin{tabular}{|ll|l|l|l|l|}
\hline
\multicolumn{2}{|l|}{Type}                                                              & Sensing head     & Maximum range & Accuracy (\%) & Bandwidth \\ \hline
\multicolumn{1}{|l|}{\multirow{3}{*}{All fiber}}         & Sagnac                 & Sagnac effect    & 5e5           & 0.1-0.5  & 1e6       \\ \cline{2-6} 
\multicolumn{1}{|l|}{}                                   & Reflective             & Faraday effect   & 5e5           & 0.1-0.5  & 1e6       \\ \cline{2-6} 
\multicolumn{1}{|l|}{}                                   & Magneto-Optical Effect & Magnet-optic     & 7e5           & 0.2-1    & 3.4e4     \\ \hline
\multicolumn{1}{|l|}{\multirow{2}{*}{Magnetostrictive }} & Interferometric fiber  & Magnetostrictive & 10            & 0.2-2    & 50        \\ \cline{2-6} 
\multicolumn{1}{|l|}{}                                   & Fiber grating          & Magnetostrictive & 1e3           & 0.2-2    & 50        \\ \hline
\multicolumn{1}{|l|}{Thermal Effect}                     & Nano fiber             & Thermal optic    & 120           & 0.2-0.5  & 500       \\ \hline
\end{tabular}
\end{scriptsize}
\end{table}

Environmental interference and defects in device structure will have an adverse effect on measurement accuracy. Therefore, overcoming the influence of environmental factors and improving the accuracy of current measurement are powerful directions for expanding large-scale applications in the future. There are the following challenges:

1) Poor environmental adaptability Although the optical current transformer has undergone theoretical simulation and simulation experiments, the environmental impact problems it faces when operating in an electrical system cannot be underestimated. The temperature difference between day and night, changes in air humidity, etc. will have a great impact on the materials used in the transformer. The temperature, humidity, and temperature difference between north and south will cause measurement data errors. Such a harsh environment will have a great impact on micro-nano optical fibers, fiber gratings, magnetic sensitive materials, etc. Improving the environmental adaptability of traditional materials or applying new materials to current transformers has become a key direction for researchers to solve complex environmental adaptation problems. 

2) DC measurement cannot operate reliably for a long time Most new current transformers use alternating magnetic fields to measure alternating currents, and sensors that can measure DC are extremely scarce. Nowadays, the construction of DC transmission lines has increased significantly, and DC transmission technology is involved in many fields. DC measurement has always been an important research direction for current transformers. There have been many years of research on DC all-optical current transformers, but in practical applications, they cannot guarantee long-term reliable operation. Their vibration resistance and resistance to external temperature interference are important factors limiting their development. Temperature-measuring fiber optic current transformers can be used not only to measure industrial frequency currents, but also to measure direct currents due to their special sensing coupling mechanism. However, due to the complexity of the field environment, it is still necessary to consider how the primary-side device can operate safely in a complex external environment. 

3) There are many measurement errors. Due to the existing sensing methods, physical quantity conversion is required, which inevitably produces measurement errors. The more conversions, the greater the system error. In essence, the way to reduce errors is to reduce the number of conversions or accurately calculate the data during the conversion process. When a sensor that requires a large number of optical devices is used, external factors such as vibration, bending, extrusion, noise, and temperature of the optical devices will affect it and cause errors. It is urgent to seek new physical sensing methods to reduce the number of conversions.

Based on the above issues, to achieve the widespread application of optical current transformers, innovations should be made in the sensing principle, sensing materials, and sensor structure of current transformers. New materials will also become the technical support for the innovation of optical current transformers. Although optical current transformers will inevitably encounter difficulties in their future development, this is exactly where we need to study and overcome them. I believe that in the near future, optical current transformers will also be one of the main equipment to be purchased in power systems.

\clearpage


\footnotesize

\bibliographystyle{ieeetr}
\bibliography{references.bib}

\end{document}